\documentclass[pre,preprint,superscriptaddress]{revtex4-2}
\usepackage{epsfig,amsmath,amsfonts,amsthm,graphicx}
\usepackage{ulem}
\usepackage{color}
\newcommand{\beq}{\begin{equation}}
\newcommand{\eeq}{\end{equation}}
\newcommand{\beqarr}{\begin{eqnarray}}
\newcommand{\eeqarr}{\end{eqnarray}}

\def\vp{\varphi}

\begin{document}

\title{Disorder fosters chimera in an array of motile particles}
\author{L. A. Smirnov}
\affiliation{Institute of Applied Physics of the Russian Academy of Sciences, Ul’yanov Str. 46, 603950, Nizhny Novgorod, Russia}
\affiliation{Department of Control Theory, Research and Education Mathematical Center ``Mathematics for Future Technologies'',
Nizhny Novgorod State University, Gagarin Av. 23, 603950, Nizhny Novgorod, Russia}
\author{M. I. Bolotov}
\affiliation{Department of Control Theory, Research and Education Mathematical Center 
``Mathematics for Future Technologies'',
Nizhny Novgorod State University, Gagarin Av. 23, 603950, Nizhny Novgorod, Russia}
\author{G. V. Osipov}
\affiliation{Department of Control Theory, Research and Education Mathematical 
Center ``Mathematics for Future Technologies'',
Nizhny Novgorod State University, Gagarin Av. 23, 603950, Nizhny Novgorod, Russia}
\author{A. Pikovsky}
\affiliation{Institute of Physics and Astronomy, Potsdam University, 14476 Potsdam-Golm, Germany}
\affiliation{Department of Control Theory, Research and Education Mathematical 
Center ``Mathematics for Future Technologies'',
Nizhny Novgorod State University, Gagarin Av. 23, 603950, Nizhny Novgorod, Russia}

\begin{abstract}
We consider an array of non-locally coupled oscillators on a ring, which for equally spaced units possesses a Kuramoto-Battogtokh
chimera regime and a synchronous state. We demonstrate that disorder in oscillators positions leads to a transition from the synchronous to the chimera state. For a static (quenched) disorder we find that the probability of synchrony survival changes, in dependence on the number of particles,  from nearly zero at small
populations to one in the thermodynamic limit. Furthermore, we demonstrate how the synchrony gets destroyed for randomly (ballistically or diffusively) moving oscillators. We show that, depending on the number of oscillators, there are different scalings of the transition time
with this number and the velocity of the units. 
\end{abstract}
\maketitle
%
%
%

\section{Introduction}
Chimera patterns, discovered by Kuramoto and Battogtokh~(KB) almost 20 years 
ago~\cite{Kuramoto-Battogtokh-02},
continue to be in focus of theoretical and experimental studies 
(see recent reviews~\cite{Panaggio-Abrams-15,Omelchenko-18}). 
Chimera is a spatial pattern in an oscillatory medium, where some subset of oscillators is 
synchronous and form
an ordered patch, while other oscillators in a disordered patch are asynchronous. There is bistability in the classical KB setup of nonlocally coupled oscillators on a ring: a chimera pattern coexists with a fully synchronized,
homogeneous in space state. In this bistable situation,
one should specially prepare initial conditions to observe chimera, because the basin of the synchronous state is rather large. Moreover, in a finite population (i.e., for a finite number of oscillators on the ring),
the synchronous state appears to be a global attractor: the chimera state is a transient, slightly irregular state, which
has a lifetime
exponentially growing with the number of oscillators~\cite{Wolfrum-Omelchenko-11}. This paper demonstrates
that disorder in the KB setup fosters the opposite: synchronous state disappears, while chimera remains stable.

The effect of disorder on chimera has been explored in several recent publications. S.~Sinha~\cite{Sinha-19}
studied different models of coupled maps and coupled oscillators, and demonstrated that with the introduction of time-varying
random links to the network of interactions, a chimera is typically destroyed, and the synchronous state establishes. 
In paper~\cite{Omelchenko_etal-15}, the effect of random links addition on the chimera state in
coupled FitzHugh-Nagumo oscillators has been studied. It has been demonstrated, that although for a small disorder, chimera survives, it becomes destroyed if the disorder is large. 

Another way to include disorder in the setup of coupled oscillators is
to assume that the units are motile particles, possibly with randomness in their 
motion. There are two ways in constructing such models: (i)
one can assume that the oscillatory dynamics of the elements
does not influence their motion, so that there is only the influence
of the positions of the units on their oscillatory dynamics (see \cite{Uriu_etal-13}),
and (ii) there is a mutual interaction between motion and internal dynamics (see, e.g., 
\cite{Hong-18,Prignano_etal-13}). For example, for locally coupled phase oscillators randomly moving on
one-dimensional lattice \cite{Uriu_etal-13}, motility
has been shown to promote a synchronous state.  For two-dimensional motions,
the authors of \cite{Beardo_etal-17}  observed that there is a resonance 
range of random velocities, for which the transition to synchrony 
is extremely slow. The authors of~\cite{Petrungaro_etal-17} explored
one-dimensional lattice with local delayed coupling, the motion of particles
was modeled by random exchanges 
of positions of nearest neighbors; in this setup, a persistent chimera was
observed in some range of parameters.
Mostly close to our setup is a recent study of Wang et al.~\cite{Wang_etal-19}.
In this work, $128$ diffusive particles on a line have been considered. Each particle
is a phase oscillator, and the coupling is nonlocal
with a cos-shaped kernel (like in the chimera studies~\cite{Abrams-Strogatz-04}).
Depending on the parameter of diffusion and coupling, both transitions from chimera 
to synchronous state and from synchronous state to chimera have been observed.
Finally, we mention an important experimental setup where moving particles
synchronize. Prindle et al.~\cite{Prindle_etal-12} realized a set of $2.5$ millions
of \textit{e.coli} bacterial cells equipped with genetically engineered
clocks, and observed their synchronization
under conditions where these cells were transported in a microfluidic device, with a 
coupling through a chemical messenger.

In this paper, we explore the effect of disorder in particles' positions
on the properties of the ``classical'' KB 
chimera~\cite{Kuramoto-Battogtokh-02}. We consider quenched disorder (random
fixed position of the particles on the ring), and dynamical disorder 
(diffusive or ballistic motion of the particles). Below we restrict
our attention to the case of slow motions, which can be explored by
comparing with the quenched case. We will show, that the number of particles
is the essential parameter governing the dynamics, and establish scaling properties
in dependence on the parameters determining the particles velocities, and on 
the number of them.

The paper is organized as follows. We introduce the model in Section~\ref{sec:bm}.
The case of quenched disorder is considered in Section~\ref{sec:qd}.
Properties of motile oscillators are considered in Section~\ref{sec:mp}. We conclude and discuss the results
in Section~\ref{sec:con}.
\section{Basic models}
\label{sec:bm}
We introduce our basic model as a generalization of the Kuramoto-Battogtokh
setup~\cite{Kuramoto-Battogtokh-02} for a 
ring of coupled phase oscillators (particles). In contradistinction to~\cite{Kuramoto-Battogtokh-02},
where equally spaced positions of the oscillators where assumed, we 
consider general positions $0\leq x_k <1$ for $N$ oscillators on the ring. 
The coupling is distance-dependent
\begin{equation}
\dot\varphi_k=\frac{1}{N}\sum_j G(x_j-x_k)\sin(\varphi_j-\varphi_k-\alpha)
\label{eq:coup}
\end{equation}
according to the kernel
\begin{equation}
G(y)=\frac{\kappa\cosh(\kappa(|y|-0.5))}{2\sinh\frac{\kappa}{2}}\;,
\label{eq:ker}
\end{equation}
which is a generalization of the exponential kernel adopted in~\cite{Kuramoto-Battogtokh-02}
to account for periodic boundary conditions on the ring. Parameter $\kappa$ determines 
the effective range of coupling; parameter $\alpha$ is the phase shift in coupling.

For positions of the particles $x_k$, we explore three models in this paper.
\begin{enumerate}
\item \textbf{Quenched disorder:} Here the positions $x_j$ 
of particles are fixed, taken 
independently from a uniform distribution on a ring. 
\item \textbf{Diffusion of the particles:} Here the particle positions 
are subject to independent white Gaussian noise terms, leading to their diffusion (with 
diffusion constant $\sigma^2$)
\begin{equation}
\dot x_j=\sigma \xi_j(t),\quad \langle \xi_j(t)\rangle=0,\;\;
\langle \xi_j(t)\xi_k(t')\rangle=\delta_{jk}\delta(t-t')\;.
\label{eq:dif}
\end{equation}
\item \textbf{Ballistic motion of the particles:} Here the particles
move with constant fixed random velocities $v_j$. 
Below we consider 
velocities as
 i.i.d. Gaussian random variables with standard deviation $\mu$.  
\end{enumerate}
In this paper we restrict ourselves to the cases of slow motion of the particles,
i.e. to the cases of small parameters $\sigma$ and $\mu$. 

\section{Quenched disorder}
\label{sec:qd}
\subsection{Observation of a transition to chimera}
We start with the case of quenched disorder. Here the only parameter is the number
of the particles $N$. At the thermodynamic limit $N\to\infty$,
one does not expect any deviation of the dynamics of disordered sets from
the dynamics of ordered configurations,
because in both cases  in the limit $N\to\infty$ one obtains a 
system of integro-differential equations
for the distribution of phases $\vp(x,t)$:
\begin{equation}
\partial_t\vp(x,t)=\int_0^1\;dy\; G(y-x)\sin(\vp(y,t)-\vp(x,t)-\alpha).
\label{eq:coupc}
\end{equation}
Population of phase oscillators~\eqref{eq:coupc}, as has been first demonstrated
by Kuramoto and Battogtokh~\cite{Kuramoto-Battogtokh-02}, possesses two attracting states:
(i) a fully synchronous state $\vp(x,t)=\psi(t)$, and (ii)
a spatially inhomogeneous chimera state with domain of synchrony (neighboring phases are closed to
each other) and asynchrony (neighboring phases are taken from a certain probability distribution).
Finite-size effects for a \textit{regular} distribution of oscillators on the ring
have been explored
by Wolfrum and Omelchenko~\cite{Wolfrum-Omelchenko-11}. The synchronous state is still stable
for any $N$, but the chimera state appeared to be a chaotic supertransient, which lives
for a time interval exponentially growing with $N$, but eventually goes into the synchronous
state.

\begin{figure}[t]
\centering
\includegraphics[width=1.0\columnwidth]{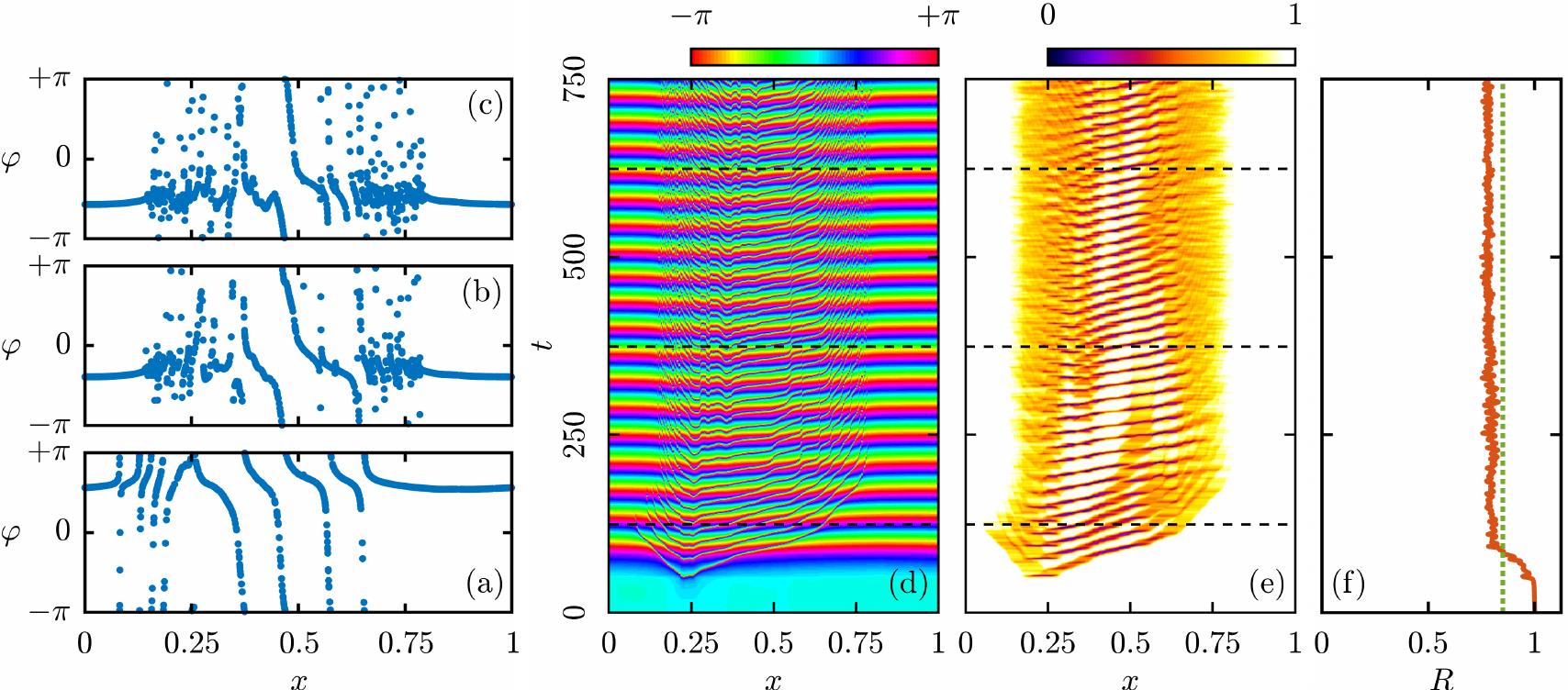}
\caption{Illustration of the transition synchrony $\to$ chimera for quenched disorder and $N=1024$ (other parameters: $\kappa=4$, $\alpha=1.457$). The particles are placed randomly on the circle, and their phases are initially equal. Panels (a, b, c): snapshots of phase distributions $\varphi(x,t)$ at (a) $t = 125$, (b) $t = 375$ and (c) $t = 625$. One can see how the synchronous state is destroyed in the presence of spatial disorder. Firstly, phase slips in a certain region of space occur. Further, clusters with the highest phase gradient begin to break down, which leads to the formation of intervals with an irregular spatial distribution of the dynamic variable $\varphi(x, t)$. After that the system goes to a chimera state. Panel (d): spatio-temporal dynamics of phases $\varphi(x,t)$. Panel (e): absolute value of the local (calculated for $K=16$ neighbors) order parameter $Z(x_n,t)=K^{-1}\sum_{k=0}^{K-1} \exp[i\vp_{n+k-K/2}]$, additionally averaged over the time interval of $3$ time units. White regions correspond to synchrony. Black dashed lines denote the moments in time for which snapshots of the phases $\varphi(x,t)$ are presented on panels (a), (b) and (c). Panel (f): the dynamics of the global order parameter $R(t)=|N^{-1}\sum_{n=0}^{N-1} \exp[i\vp_{n}]|$. It is clearly seen how the transition from the initially synchronous regime with $R=1$ to the chimera state with $R \approx 0.79$ occurs. The green dashed line shows the value $R=0.85$, which is further taken as a criterion that determines the time of destruction of the synchronous mode.}
\label{fig:ill1}
\end{figure}

Our main observation is that the opposite happens for an \textit{irregular} 
distribution
of oscillators on the ring. Namely, the initial synchronous 
state may become destroyed
for finite $N$, while the chimera state is stable. 
We illustrate a transition from the synchronous to chimera regime
in Fig.~\ref{fig:ill1}.

Qualitatively, destruction of the synchronous state due to disorder is similar to desynchronization
in disordered oscillator lattices first described by  
Ermentrout and Kopell~\cite{Ermentrout-Kopell-84}. At large enough disorder a synchronous state
in the lattice disappears due to a saddle-node bifurcation. In our setup we cannot directly apply
theory~\cite{Ermentrout-Kopell-84}, because we have a ring with long-range coupling. Furthermore,
theory~\cite{Ermentrout-Kopell-84} is restricted to the case $\alpha=0$, while
in our setup parameter $\alpha$ is close to $\pi/2$.

\subsection{Statistical evaluation}
\begin{figure}[t]
\centering
\includegraphics[width=0.7\columnwidth]{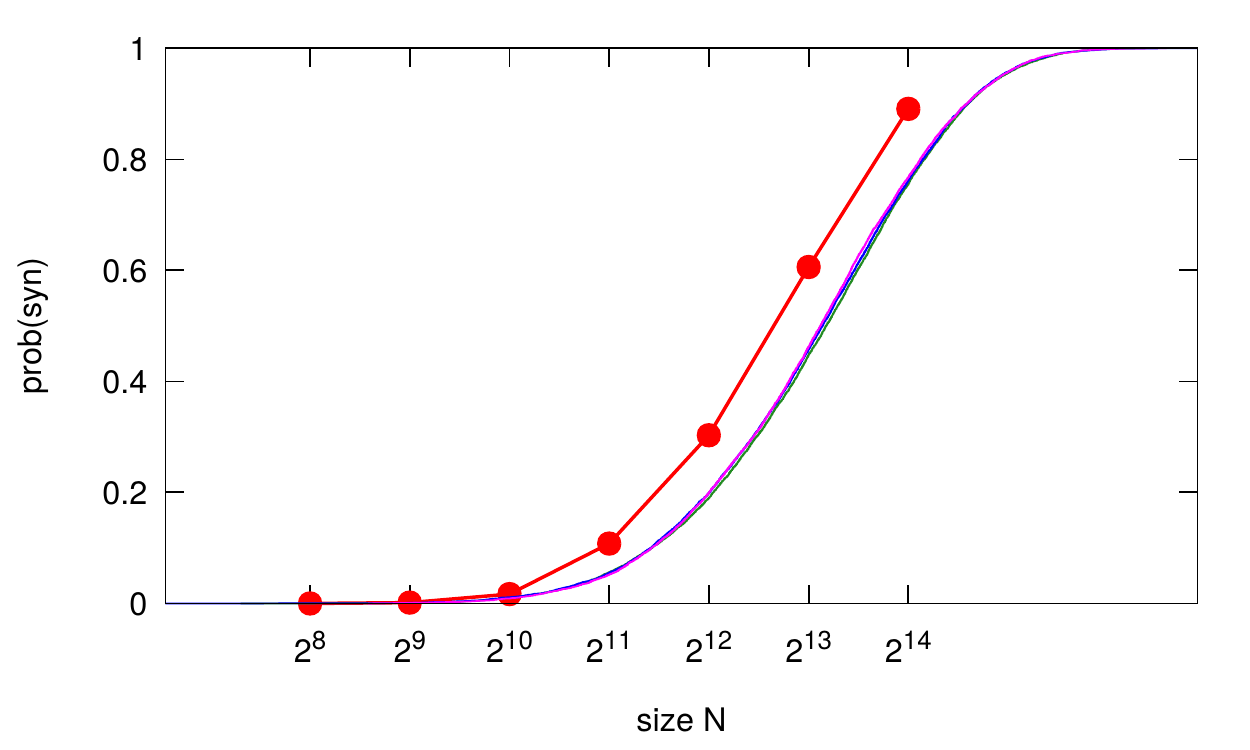}
\caption{Red dots: probabilities of existence of a synchronous state from direct numerical simulations.
Curves: rescaled cumulative distributions of the maximum of field $H$, for $N=128$ (green curve),
$N=256$ (blue) and $N=512$ (magenta). These curves are drawn with help of expression
\eqref{eq:anes} and  practically overlap, what confirms the validity of the scaling $\sim N^{1/2}$.}
\label{fig:static}
\end{figure}

In Figure~\ref{fig:static} we present a direct statistical evaluation of the 
probability for synchrony
to occur. The numerical experiment 
has been performed as follows:
for a configuration of random positions of oscillators $x_j$, equations \eqref{eq:coup}
were solved starting from the state with all phases being equal $\vp_1=\ldots=\vp_N$.
If a steady rotating state where all the instantaneous frequencies are equal
appears, the  configuration is considered as a synchronous one.
Otherwise, if in the set of oscillators phase slips appear, the configuration
is considered as a non-synchronous (chimera). Many runs with
random positions have been sampled to achieve statistical results presented in the Fig.~\ref{fig:static}.
One can see that while the probability to observe synchrony is very low for
relatively small $N$ (in fact, for $N=256$ no any synchronous case out of $10^4$ runs has been observed),
it becomes high for $N\gtrsim 8192$. This confirms the qualitative picture of the local
stability of the synchronous state at $N\to\infty$. We stress here that we do not
consider here very small systems with a few oscillators.

\subsection{Analytic estimate of  probability of the existence of stable synchrony}

Here we give a semi-analytic estimate for the probability to observe a 
synchronous state in a disordered array. Instead of performing a rigorous
bifurcation analysis, we first estimate (approximately), at which fluctuation of the
local acting field, the synchronous state disappears. If we assume that
in the synchronous state all the phases are equal and rotate with
the frequency $\Omega=-\sin\alpha$ (in fact, this is only true for the 
regular uniform distribution of the units), then from Eq.~\eqref{eq:coup}
it follows that an oscillator $\varphi_k$ will not be able to follow
this collective synchrony if the acting field on it $H(x_k)=N^{-1}\sum G(x_j-x_k)$
exceeds the threshold $H_c=2-\sin\alpha$. Thus, to find the probability that this happens, we
have to analyze the distribution of maxima of the field $H(x)$ defined as
\begin{equation}
H(x)=\frac{1}{N}\sum_{k=1}^{N}G(x-x_k)\;,
\label{eq:af}
\end{equation}
where $x_k$ are random positions on the interval $0\leq x<1$ with
uniform density $w(x)=1$. Statistics of the field $H$ can be evaluated as follows.
First, due to normalization $\int_0^1G(y)dy=1$, we get $\langle H\rangle=1$.
Next, using independence of positions $x_k$, it 
is straightforward to calculate the covariance of $H$
(this calculation is completely analogous to a calculation of the correlation 
function of the shot noise (sequence of independent pulses, the Campbell's formula) \cite{Beichelt-06}):
\begin{equation}
\begin{aligned}
K(y)=&\langle H(x)H(x+y)\rangle-1=N^{-1}\frac{\kappa^2 B(\kappa,y)}{4\sinh^2 
\frac{\kappa}{2}}\;,\\
B(\kappa,y)=&\frac{\cosh \kappa y}{2}+\frac{y[\cosh \kappa(y-1)-\cosh\kappa y]}{2}
+\frac{\sinh \kappa y -\sinh\kappa(y-1)}{\kappa}\;.
\end{aligned}
\label{eq:corr}
\end{equation}
One can see that the variance of field $H$ decays as expected $\sim N^{-1}$.
One can argue that for large $N$, as a sum of $N$
statistically independent contributions, the field $H(x)$ is Gaussian, and
this indeed is nicely confirmed by numerics (not shown). However, we
are interested in the distribution of the \textit{maximum} if this field,
and obtaining it is a nontrivial task, because of correlations \eqref{eq:corr}
(cf.~\cite{Nadarajah_etal-19}).
These correlations, however, do not depend on $N$ except for a factor $N^{-1}$,
and therefore one can expect that the scaled 
distribution of the maximum $h=(H_{max}-1)\sqrt{N}$ will be system-size-independent.
The cumulative distribution function of maxima $W(h)$ thus provides
an estimate that for an ensemble of size $N$ the synchronous population
survives: 
\begin{equation}
P_s(N)=W((N)^{1/2}(1-\sin\alpha)). 
\label{eq:anes}
\end{equation}
In Fig.~\ref{fig:static}
we compare this estimate with direct numerical simulations, using three
distributions $W$ obtained for $N=128,256,512$. These curves are
practically indistinguishable, what is just another manifestation of validity of
the scaling $H_{max}-1\sim N^{-1/2}$. The curve lies below the numerical data,
 what means that
the adopted estimate is rather crude. Nevertheless, it correctly predicts that for $N\lesssim 1000$
practically all configurations lead to a chimera state.

\section{Transition from synchrony to chimera for motile particles}
\label{sec:mp}

\begin{figure}[t]
\centering
\includegraphics[width=1.0\columnwidth]{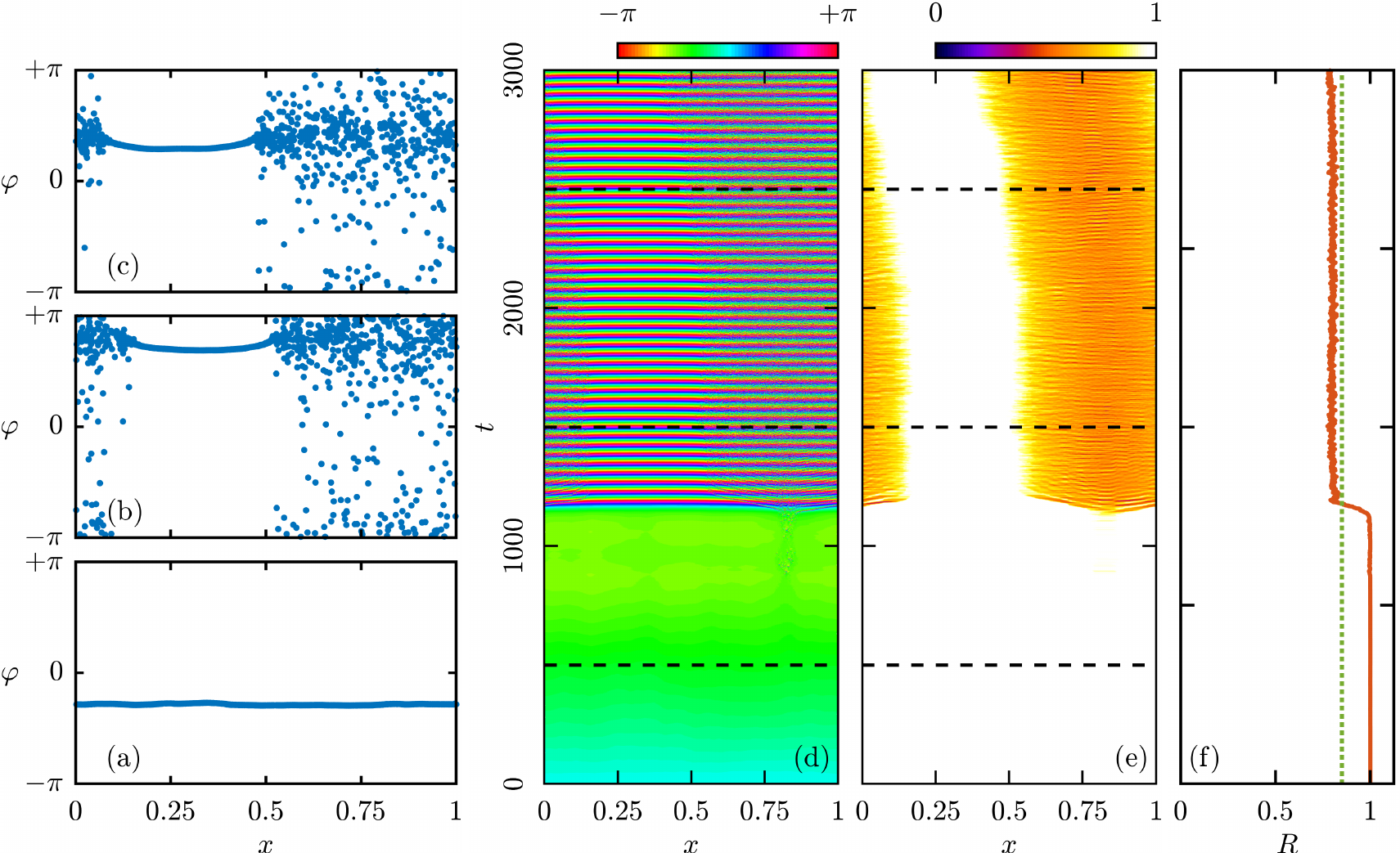}
\caption{The same as Fig.~\ref{fig:ill1}, but for diffusive particles with $\sigma=10^{-3}$. 
Initially all the particles are placed equidistantly on the circle, and have equal phases. Developing at $t\approx 1000$ chimera pattern slowly moves along the circle, due to random rearrangement of particles positions. Panels (a, b, c): snapshots of phase distributions $\varphi(x,t)$ at (a) $t = 500$, (b) $t = 1500$, (c) $t = 2500$. Panel (d): spatio-temporal dynamics of phases $\varphi(x,t)$. Panel (e): absolute value of the local order parameter $Z(x,t)$. Panel (f): the dynamics of the global order parameter $R(t)$.}
\label{fig:ill2}
\end{figure}

In this section we consider motile particles with random trajectories.
In all cases reported in this section below, we start at $t=0$ with particles
regularly distributed over the ring, i.e. $x_k(0)=(k-1)/N$. The phases
are set to be equal, so that the initial state is the perfectly synchronized one.
Because of irregular motion, disorder in the position of the particles appears. 
At rather large times the particles can be considered as noncorrelated, thus
their positions are fully random on the ring. This, as we have seen in Section~\ref{sec:qd},
facilitates transition to chimera. Moreover, as in the course of time evolution 
different random configurations appear, eventually one which does not support
synchrony will lead to a transition to chimera (we illustrate this
in Fig.~\ref{fig:ill2}). Thus, on the contrary to the case
of static configurations of Section~\ref{sec:qd}, we expect that a transition from synchrony to chimera will always be observed even at system sizes as large as $N=8192$.

Our main interest below is in the dependence of the transition time (from synchrony to chimera) 
on the parameters
of noise and particle size. 
In the system of differential equation~\eqref{eq:coup}, positions of the particles $x_k$
can be considered as parameters. We start with a stable fixed point in this system, which does exist
for regularly spread particles. Slow motion of particles means slow variation
of the parameters in \eqref{eq:coup}, and initially the stable steady state continues to exist.
However, when the set of parameters reaches a bifurcation point (numerical experiments show that
this is a saddle-node bifurcation, like in a disordered lattice~\cite{Ermentrout-Kopell-84}),
the steady state disappears and another, chimera state, appears. Thus, what we want to study, 
is the time
to bifurcation. 

There is also another view on the transition to chimera. In the starting configuration, where
the oscillators are equidistantly distributed, the acting field $H(x)$ (see Eq.~\eqref{eq:af})
is constant. When the particles start moving, this field is no more constant, so one observes
roughening of $H(x)$~\cite{Barabasi-Stanley-95}. This roughening continues until the maximum of the 
field becomes large enough to produce the bifurcation. This picture suggests that
one can expect the average time of the transition $\langle T\rangle$ to scale with parameters
of the problem: characteristic random velocities of the particles and the number of them. We
explore this idea of scaling below.

\begin{figure}[!htb]
\centering
\includegraphics[width=0.48\columnwidth]{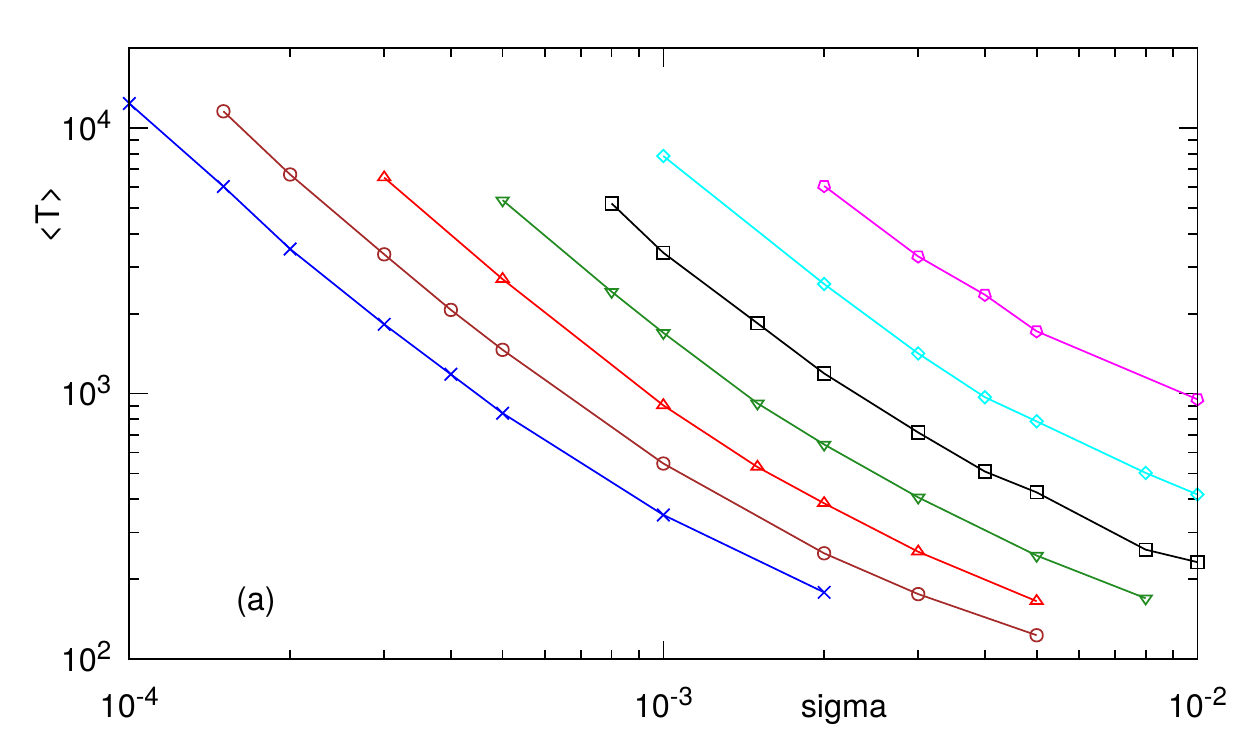}\hfill
\includegraphics[width=0.48\columnwidth]{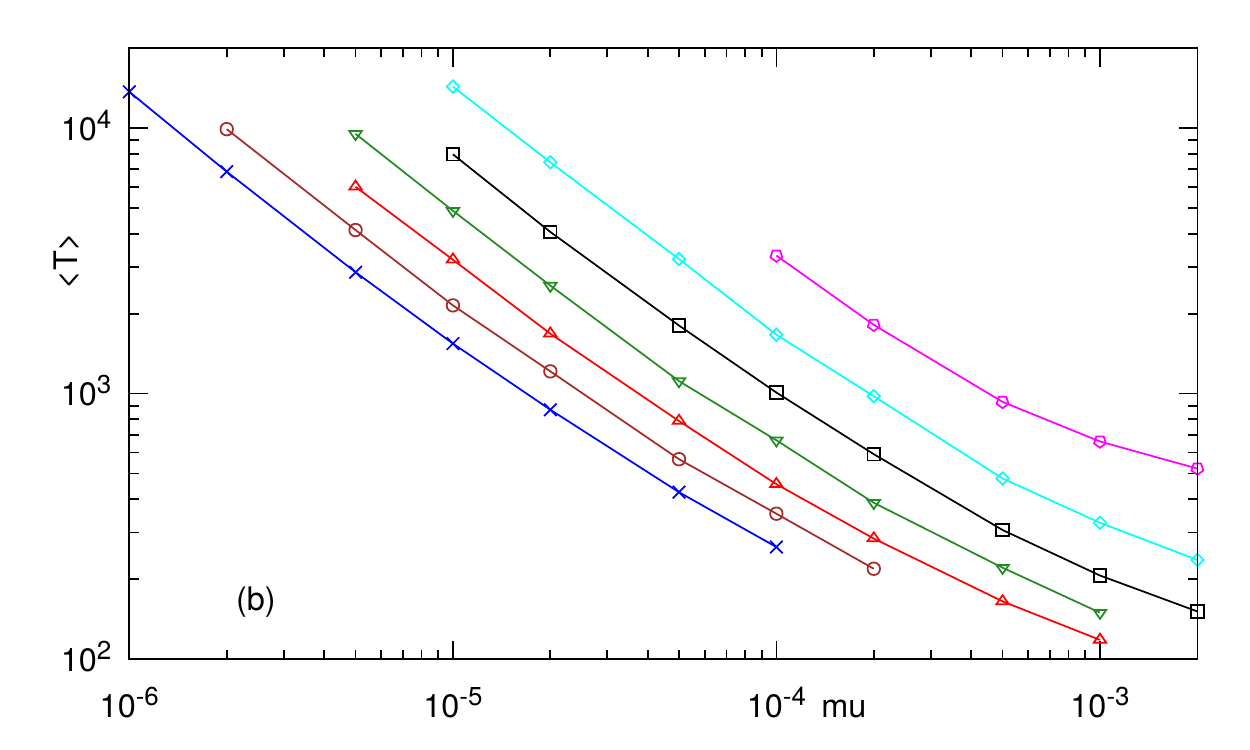}
\caption{Average time for a transition from synchrony to chimera for different
$N$ (from $N=128$ (bottom curve) to $N=8192$ (top curve), values of $N$ increase by factor $2$;
markers and colors are the same as in Fig.~\ref{fig:scal}).
Panel (a): 
for diffusive motion of the particles \eqref{eq:dif} in dependence on the diffusion parameter $\sigma$. 
Panel (b): for ballistic particles 
with Gaussian distribution of velocities, in dependence on the standard deviation $\mu$. }
\label{fig:tim}
\end{figure}

We consider two basic setups for the random motion of particles:
\begin{enumerate}
\item{\bf Diffusive motion.}
Here we consider diffusive motion of the particles according to \eqref{eq:dif}.
The average transition times from synchrony to chimera are presented in Fig.~\ref{fig:tim}(a). 
As expected, the time grows with the number
of particles $N$, and for small diffusion rates $\sigma$. 
\item{\bf Ballistic motion.}
Here we assume that the particles move with constant velocities $v_j$,
which are chosen from the normal distribution with standard variation $\mu$. The average 
transition times are shown in Fig.~\ref{fig:tim}(b).
\end{enumerate}

\begin{figure}[t]
\centering
\includegraphics[width=0.49\columnwidth]{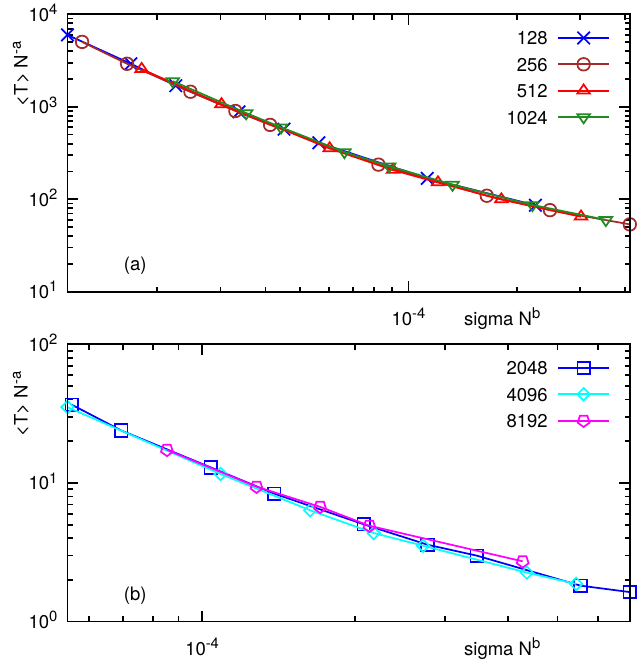}\hfill
\includegraphics[width=0.49\columnwidth]{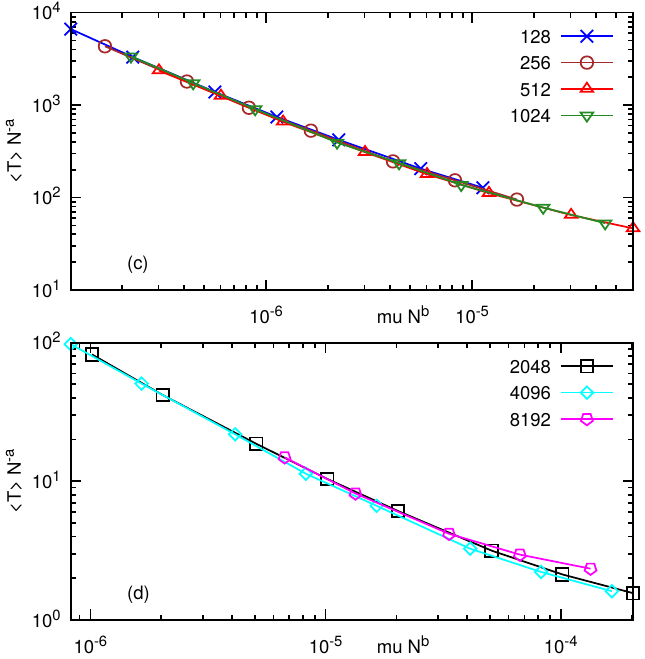}
\caption{ The same data as in Fig.~\ref{fig:tim}, but in scaled coordinates
(diffusive particles in panels (a,b), ballistic particles in panels (c,d)).
Top row (panels (a,c)): scaling for $128\leq N\leq 1024$ with $b=0.45$ and $a=0.15$.
Bottom row (panels (b,d)): scaling for $2048\leq N\leq 8192$ with $b=0.35$ and $a=0.65$
for diffusive particles, and $b=0.3$ and $a=0.6$ for ballistic particles.}
\label{fig:scal}
\end{figure}

Next, we discuss scaling properties of the time to chimera. We look for the scaling 
relation in the form
\begin{equation}
\langle T(c,N)\rangle=N^{a}f\left(\frac{c}{N^b}\right)
\label{eq:scr}
\end{equation}
where $c$ stays for one of the parameters $\mu,\sigma$, and constants $a,b$ generally
depend on the setup. We, however, could not fit all the data according to a unique law
\eqref{eq:scr}. As we illustrate in Fig.~\ref{fig:scal}, taking data for 
the interval of system sizes 
$128\leq N\leq 1024$ allows to achieve a very good collapse
of data points using scaling in form \eqref{eq:scr},
with $b=0.45$ and $a=0.15$ for both cases (diffusive and ballistic motions). However,
using these parameters for larger system sizes $N\geq 2048$ does not lead to a
good collapse of points. Rather we use for large $N$ values $b=0.3$ and $a=0.6$
for the ballistic case and  $b=0.35$ and $a=0.65$ for the diffusive case,
but they result only in an approximate collapse of data points. 

We attribute this absence of a universal scaling to the properties of
the quenched randomness described in section~\ref{sec:qd}. As it follows
from Figure~\ref{fig:static}, for $N\lesssim 1024$ it is enough for particles 
to achieve random independent positions on the circle, then the transition to chimera is 
nearly certain. In contradistinction, for larger populations there is a finite
probability for a random quenched configuration to possess synchrony. This leads to an increase
of the transition time: random motion of the particles explores different
configurations, until one that does not posses synchrony is found and the transition
to chimera occurs. This explains different scalings with a crossover near $N=1024$.
Moreover, we expect that the scaling observed for $2048\leq N\leq 8192$
will not extend to larger system sizes, because according
to Figure~\ref{fig:static}, the probability of the transition in quenched configuration 
drastically reduces, so that the time to achieve chimera will be extremely large, if not infinite.

\section{Conclusion}
\label{sec:con}

In this paper we studied the effect of the oscillators position disorder on the
chimera state in the Kuramoto-Battogtokh model of nonlocally coupled phase
oscillators on a ring. The level of disorder is basically determined
by the number of units $N$, it disappears in the thermodynamic
limit $N\to\infty$. Our main finding is that large 
disorder facilitates stability
of chimera, and for sizes of populations below some level, it is practically
impossible to observe  a stable synchronous regime in a setup with a
quenched disorder. For slow random motions of the particles, in the explored range
of system sizes up to $N=8192$, we observed a transition from synchronous 
initial configuration to a chimera in all realizations. Even when synchrony has
a finite probability to exist in a quenched configuration, slow variations 
of positions of particles lead eventually to a configuration where synchrony state
does not exist, so that a chimera develops. 

We explored the scaling properties
of the transition to chimera and found that for both diffusive and ballistic motions,
the scaling exponents in the relation~\eqref{eq:scr} are nearly the same. Due to a nontrivial
dependence of the probability of the existence of synchrony already for quenched disorder,
the scaling is different for relatively small sizes $N$ (where synchrony is
practically never observed
) and for larger sizes, where in the quenched case there is a finite probability for synchrony
to survive. We, however, have not explored very large populations $N > 8192$,
because of computational restrictions.

We stress here that we studied the Kuramoto-Battogtokh model for the ``standard''
parameters $\kappa,\alpha$ used in \cite{Kuramoto-Battogtokh-02}. The domain
of existence of chimera and its basin of attraction may depend significantly
on these parameters. Extension of the obtained results on other domains of 
parameters and on other setups where chimera patterns exist is a subject of ongoing study.

In this paper we focused on the regime of very slow motion of the particles,
including the static (quenched) case. Preliminary simulations show that the regimes with fast 
particles can differ significantly, this is a subject of ongoing research.
Another interesting case for future exploration is one close to the thermodynamic limit,
where finite-size fluctuations are small. Here an analytical description
based on the Ott-Antonsen reduction might be possible, to be reported elsewhere.

\begin{acknowledgments}
We thank O. Omelchenko for fruitful discussions. A. P.
acknowledges support by the Russian Science Foundation 
(grant Nr.~19-12-00367) and by DFG (grant PI 220/22-1).
\end{acknowledgments}
\def\cprime{$'$}

\end{document}